\RequirePackage{ifpdf}
\ifpdf 
\documentclass[pdftex]{sigma}
\else
\documentclass{sigma}
\fi

\numberwithin{equation}{section}

\begin{document}

\allowdisplaybreaks

\renewcommand{\thefootnote}{$\star$}

\renewcommand{\PaperNumber}{044}

\FirstPageHeading

\ShortArticleName{Semi-Harmonic Rectangular Potentials}

\ArticleName{Rectangular Potentials in a Semi-Harmonic\\ Background:
Spectrum, Resonances and Dwell Time\footnote{This
paper is a contribution to the Proceedings of the Workshop ``Supersymmetric Quantum Mechanics and Spectral Design'' (July 18--30, 2010, Benasque, Spain). The full collection
is available at
\href{http://www.emis.de/journals/SIGMA/SUSYQM2010.html}{http://www.emis.de/journals/SIGMA/SUSYQM2010.html}}}

\Author{Nicol\'as FERN\'ANDEZ-GARC\'{I}A~$^\dag$ and Oscar ROSAS-ORTIZ~$^\ddag$}

\AuthorNameForHeading{N.~Fern\'andez-Garc\'{\i}a and O.~Rosas-Ortiz}

\Address{$^\dag$~Instituto de F\'{\i}sica, UNAM, AP 20-353, 01000
M\'exico D.F., Mexico}
\EmailD{\href{mailto:nicolas@fisica.unam.mx}{nicolas@fisica.unam.mx}}

\Address{$^\ddag$~Physics Department, Cinvestav, A.P. 14-740,
M\'exico DF 07000, Mexico}
\EmailD{\href{mailto:orosas@fis.cinvestav.mx}{orosas@fis.cinvestav.mx}}

\ArticleDates{Received December 01, 2010, in f\/inal form April 29, 2011;  Published online May 05, 2011}

\Abstract{We study the energy properties of a particle in one
dimensional semi-harmonic rectangular wells and barriers. The
integration of energies is obtained by solving a simple
transcendental equation. Scattering states are shown to include
cases in which the impinging particle is `captured' by the
semi-harmonic rectangular potentials. The `time of capture' is
connected with the dwell time of the scattered wave. Using the
particle absorption method, it is shown that the dwell time
$\tau^a_D$ coincides with the phase time $\tau_W$ of Eisenbud and
Wigner, calculated as the energy derivative of the ref\/lected wave
phase shift. Analytical expressions are derived for the phase time
$\tau_W$ of the semi-harmonic delta well and barrier potentials.}

\Keywords{exactly solvable potentials; scattering process;
resonances; Eisenbud--Wigner phase time; dwell time}

\Classification{35Q40; 35B34; 81U30; 81Q60}

\renewcommand{\thefootnote}{\arabic{footnote}}
\setcounter{footnote}{0}

\section{Introduction}

One-dimensional models of quantum mechanics are useful in a number
of applications in contemporary physics
\cite{Bas82,Cap92,Bra89,Ros08,Rot09,Gar10,Fer10,Esp08,Lud87,Emm05,Ant01,
del02,del03,Zav04,Fer08b,Kla10,Chr03,Flu99}. Often employed into
the approximations which make tractable the more elaborated
three-dimensional systems, they allow to get a deeper insight on
the physics involved. Their simplicity has made them valuable as
academic and research tools. For instance, the concept of
ef\/fective mass, successfully applied in describing the formation
of shallow energy levels due to impurities in crystals, leads to
one-dimensional systems \cite{Bas82}. Localized lattice
deformations can be modelled as a rectangular potential to study
isolated transitions, observed in semiconductors, from a bound
state within a quantum well to a bound state at an energy greater
than the barrier height \cite{Cap92}. The analysis of resonances
\cite{Bra89,Ros08,Rot09,Gar10,Fer10,Esp08} is transparent for
rectangular potentials in either, the presence of a background
interaction \cite{Lud87,Emm05}, or in free space
\cite{Ant01,del02,del03,Zav04,Fer08b,Kla10,Chr03}. Simple mo\-dels
of point-like \cite{Flu99,Gad10}, as well as regularized singular
interactions \cite{Sca58}, can be obtained as limit cases of
rectangular potentials \cite{Neg02,Chr03,Neg04}. The
one-dimensional models are also useful in the study of
supersymmetric quantum mechanics
\cite{Mie04,Kha04,Bay04,And04,Ferd10}. Based on the Darboux
transformations \cite{And84} (see also \cite{Mie04}), the
supersymmetric (intertwining or factorization) formalism allows
the construction of new exactly solvable potentials even if
complex energies are involved
\cite{Bay96,And99,Fer03,Ros03a,Ros07}. This last property has been
implemented to get one-dimensional complex potentials behaving as
an optical device which both refracts and absorbs light waves
\cite{Fer08b,Ros07}.

\looseness=-1
Collisions, on the other hand, are modelled as interactions
localized in time and space. This implies that the involved
potential vanishes rapidly enough in space, so that incoming and
outgoing asymptotic states can be represented by wave packets in
free motion. The problem is usually reduced to the analysis of a
one-dimensional ef\/fective potential. Main information is then
obtained from the transmission and ref\/lection amplitudes. Of particular interest, the
resonance phenomenon is experimentally studied in atomic, nuclear
and particle physics \cite{Bra89,Ros08,Rot09,Gar10}. A resonance can be understood
as an special result of the scattering process in which the
incident wave is `captured' by the scatterer for a while. A
measurable {\em time delay} (relative to free motion) is then
associated to the scattered wave~\cite{Tay06,Boh51}.

The concept of time delay corresponds to the time spent by an
scattered particle in the scattering zone when compared to a free
particle subject to the same initial conditions. According to
Eisenbud and Wigner, this is associated to the energy derivative
of the phase shift for binary collisions (see~\cite{Wig55,Smi60}). The
existence of a global time delay (as connected to sojourn times),
and its identity with the Eisenbud--Wigner phase time, have been
proved for local potentials in~$\mathbb R^3$~\cite{Amr87}. The proof has been extended to an
abstract formalism where time delay is def\/ined in terms of the
expectation values of non-negative normalized functions of compact
support~\cite{Ric10}. In general, the retardation of the scattered wave
involves transient ef\/fects which are relevant in nuclear reactions
\cite{Mos51,Mos52}, and is also fundamental in the
characterization of resonances \cite{Tay06,Boh51}. The time delay is also connected with the Levinson theorem~\cite{Osb77}, the density of states in mesoscopic conductors
\cite{Ped98} and with the study of the photodetachment rate due to
weak time-periodic electric f\/ields \cite{Emm00}. In
one-dimensional systems, the time delay has been studied for
step-like potentials exhibiting a two-channel structure
\cite{Amr07}. Quite recently, it has been shown that time delay
can be obtained from the eigenvalues of a non-Hermitian
Hamiltonian in two and three dimensions \cite{Bar10}. Extensive,
clear reviews with complete bibliography can be found in~\cite{Hau89,Car02,Sas10}.

\looseness=-1
The present work is addressed to the analysis of the energy
properties of a particle subject to the action of a semi-harmonic
rectangular potential. This last is either a rectangular well or
barrier in a background integrated by a free-particle interaction
to the right and an oscillator-like interaction to the left of the
rectangular potential. The model corresponds to a system (the
rectangular potential) embedded in an environment (the
semi-harmonic background), and the issue is the study of the
modif\/ications on the energy spectrum and resonances of the system
which are induced by the environment. Thus, the semi-harmonic
square potentials are viewed as open one-dimensional quantum
systems \cite{Rot09}. The bound states and resonances of square
potentials have been studied by using diverse approaches (see,
e.g., \cite{Zav04,Fer08b,Kla10} and references quoted therein). In
particular, if closer resonances imply narrower widths (i.e., in
the case of isolated resonances), the transmission amplitude $T$
can be written as a superposition of Fock--Breit--Wigner
distributions~\cite{Fer08b}. Then, the position $E$ and width
$\Gamma$ of a resonance $\epsilon = E - \frac{i}{2} \Gamma$ are in
correspondence with one of the bell-shaped peaks of $T$. This
result is extended to the case of resonances which are not
isolated by identifying the position of the peaks with the
absolute value of $\epsilon$, rather than using
$\mbox{Re}(\epsilon) =E$~\cite{Kla10}. In the limit where the
square potentials become a point-like interaction
(i.e., a delta barrier or well), the transparency properties
involved produce no resonances~\cite{Flu99}. The situation is dif\/ferent
for dipole-like interactions represented by derivatives of the
delta distribution with respect to the position~\cite{Chr03}.

As open systems, rectangular potentials
have been previously studied in a
static f\/ield as the environment~\cite{Lud87,Emm05}; the reports
include the point-like limit~\cite{Emm00}. Unlike the rectangular
potentials in the free-particle background, these last open
systems are not isotropic since the environment is represented by
a potential diverging as~$\pm x$ at~$\vert x \vert = + \infty$.
Thus, though the presence of resonances is ensured by adding such
a~background, the conventional def\/initions of dwell time and time
delay are not automatically applicable in these cases. Indeed,
most of the approaches on the matter consider a potential which is
very localized in a f\/inite region of space
\cite{Hau89,Car02,Sas10} (see also \cite{Amr87}). In
contraposition, the open systems presented in this paper become
zero when they are evaluated at positions to the right of the
square potential. This condition represents an advantage over the
square potentials in a constant f\/ield environment since ref\/lection
times can be now calculated. In this context, the mean time spent
by the particle in the interaction zone is def\/ined as the dwell
time (or sojourn time) involved. However, it must be emphasized
that the parabolic part of the environment makes not simple the
calculation of transmission times. Thereby, there is not a clear
def\/inition of time delay in terms of dwell time dif\/ferences in
arbitrary open balls of the one-dimensional domain of these
systems. Instead, the time delay can be def\/ined as the dif\/ference
between the ref\/lection times above mentioned. Here, we shall show
that the dwell time coincides with the phase time introduced
by Eisenbud and Wigner.

The organization of the paper is as follows. In
Section~\ref{sec:2} we establish the general expressions
concerning the semi-harmonic rectangular potentials. The condition
for the trapping of particles is identif\/ied as the f\/inding of
zeros of the Jost function. A very simple transcendental equation
is derived for the numerical integration of energies. In
Section~\ref{sec:3}, the family of semi-harmonic rectangular wells
of unit area is shown to converge to a semi-harmonic delta well in
the sense of distribution theory. The corresponding transcendental
equation is shown to admit a unique root associated to bound
states, and the result is compared with the single bound energy of
the delta well in a free particle background. The resonances obey
a rule of distribution in the complex plane of the energies that
resembles the distribution of the odd bound energies of a harmonic
oscillator. Section~\ref{sec:4} is devoted to the application of
the previous results to the case of a semi-harmonic rectangular
barrier. Interestingly, the limit case of a semi-harmonic delta
barrier obeys a distribution of resonances which is in
correspondence with the even energy eigenvalues of the harmonic
oscillator. A brief discussion on the Darboux transformations of
rectangular potentials in a semi-harmonic background is also
given. In Section~\ref{sec:5} we investigate the dwell time of the
scattering process associated to the semi-harmonic rectangular
potentials. This is def\/ined as the time spent by the incident
particle in the interaction zone. We use the particle absorption
method \cite{Gol90,Hua91,Mug92} to determine the characteristic
time involved. Then we assume that the probability of f\/inding the
particle decays exponentially from the moment that it is
`captured' by the potential. The dwell time so calculated
coincides with the phase time~$\tau_W$ of Eisenbud and Wigner. For
delta wells and barriers explicit analytical expressions of~$\tau_W$ are given. In each case, the time delay $\tau_W$ has
local maxima centered at the real part of the involved resonances.
This last, combined with the results of Sections~\ref{sec:3} and~\ref{sec:4} for delta-like potentials, indicates that the
semi-harmonic background induces the real part of the resonances
to be distributed in correspondence with the energy eigenvalues of
the harmonic oscillator. Some concluding remarks are given in
Section~\ref{sec:6}. Finally, a short appendix includes some
derivations in terms of the conf\/luent hypergeometric functions
which, although important, can be postponed to a later reading.

\section{Semi-harmonic rectangular well}
\label{sec:2}

\begin{figure}[t]
\centering\includegraphics[width=6cm]{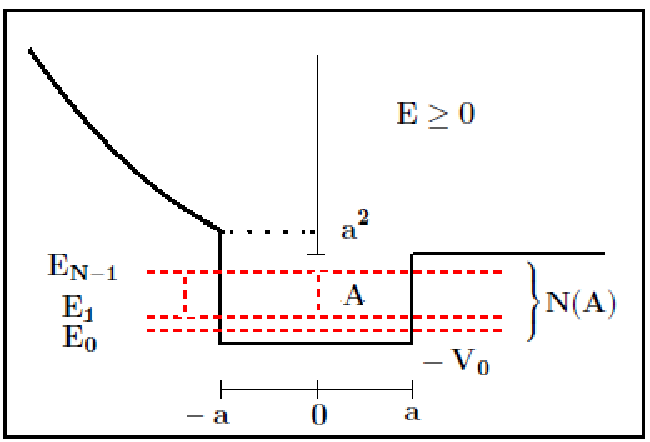}

\caption{The rectangular well in a semi-harmonic background. The
number $N$ of bound state energies $E_0, \ldots, E_{N-1}$, depends
on the area $A=2aV_0$ of the well (cf.\ Section~\ref{bound}). Some of
the incoming scattering waves of energy $E\geq 0$ are `captured'
for a while in the region $(-\infty, a)$, as discussed in
Section~\ref{sec:5}. If $V_0^{-1}=2a$, the limit $a\rightarrow 0$
leads to a semi-harmonic delta well (see Section~\ref{sec:3}).
Semi-harmonic rectangular and delta barriers are obtained by the
change $V_0 \rightarrow -V_0$ (cf.\ Section~\ref{sec:4}).}
\label{Fig:pozo}
\end{figure}

Consider a particle of energy $E$ which is under the inf\/luence of
the one-dimensional potential
\begin{gather}
V(x; a)=\left\{
\begin{array}{rl}
x^2, & x \leq -a \quad (\mbox{\rm Region I}),\\[1ex]
-V_0, & \vert x \vert < a \quad (\mbox{\rm Region II}),\\[1ex]
0, & a \leq x \quad (\mbox{\rm Region III}),
\end{array}
\right.  \qquad a \geq 0, \qquad V_0 > 0.
\label{pot}
\end{gather}
Potential (\ref{pot}) is depicted in Fig.~\ref{Fig:pozo}. The
corresponding stationary dimensionless Schr\"odinger equation
\begin{gather}
(H-E)\psi(x)=0, \qquad H:= -\frac{d^2}{dx^2}+V(x,a)
\label{schro}
\end{gather}
is solved by the appropriate combination of the functions
\begin{gather*}
\psi_{\rm I}(x)= A_1 e^{-x^2/2}\, {}_1F_1\big(\alpha, \tfrac12 ;x^2\big) + B_1
e^{-x^2/2} x \, {}_1F_1\big(\alpha +\tfrac12, \tfrac{3}{2} ;x^2\big),\nonumber\\
\psi_{\rm II}(x) = A_2 u(x)+B_2 v(x):= A_2 \sin qx + B_2 \cos qx,\\ 
\psi_{\rm III}(x)= A_3 e^{ikx}+ B_3e^{-ikx},\nonumber
\end{gather*}
where $k= \sqrt E$ and $q = \sqrt{V_0+k^2}$, the functions
$e^{ikx}$ and $e^{-ikx}$ represent waves moving towards the right
and the left respectively, and ${}_1F_1(\alpha,\gamma;z)$ stands
for the conf\/luent hypergeometric function (see Appendix) with
$\alpha =\frac{1-k^2}{4}$.

To analyze the solutions of (\ref{schro}) as $x \rightarrow
-\infty$, we use (\ref{2asymp-}). A simple calculation gives
\begin{gather*}
\psi_{\rm I}(x) \approx \frac{\sqrt \pi}{\Gamma_{\mbox{e}} (\alpha)}
e^{\frac{x^2}{2}} (x e^{-i\pi})^{2\alpha-1} \left[ A_1
-\frac{B_1}{2} \frac{\Gamma_{\mbox{e}} (\alpha)}{\Gamma_{\mbox{e}}
(\alpha +1/2)} \right], \qquad \mbox{\rm as} \quad x\rightarrow
-\infty,
\end{gather*}
with $\Gamma_{\mbox{e}}(z)$ the Euler gamma function of $z\in
\mathbb C$. Therefore, to construct solutions which are regular at
the left edge of $\mbox{\rm Dom} \,V(x,a)=(-\infty,\infty)$, the
coef\/f\/icient $B_1$ is constrained to satisfy
\begin{gather*}
B_1= 2A_1\frac{\Gamma_{\mbox{e}} (\alpha +1/2)}{\Gamma_{\mbox{e}}
(\alpha)}.
\end{gather*}
As a consequence, the function $\psi_{\rm I}$ reads
\begin{gather}
\psi_{\rm I}(x)=A_1 e^{-x^2/2}\left[ {}_1F_1\big(\alpha, \tfrac12;x^2\big)+ 2x
\frac{\Gamma_{\mbox{e}} (\alpha +1/2)}{\Gamma_{\mbox{e}} (\alpha)}
\, {}_1F_1\big(\alpha +\tfrac12,\tfrac{3}{2};x^2\big)\right]:= A_1 \varphi(x),
\label{psi1}
\end{gather}
and it becomes zero as $x\rightarrow -\infty$. If the parabolic
part of the potential appears to the right, rather than to the
left of the well, it occurs a phase dif\/ference in the asymptotic
behavior of the solution (compare equations \eqref{2asymp+} and
\eqref{2asymp-}). We decided to put the oscillator-like interaction
to the left of the well to precisely emphasize the subtleties of
the involved mathematics.

The condition of continuity at $x=-a$ leads to the coef\/f\/icients
\begin{gather*}
\frac{A_2}{A_1}= \left. \frac{\varphi v (\beta_v -
\beta_{\varphi})}{q}\right\vert_{x=-a}:=\varphi(-a) C_2,
\qquad \left. \frac{B_2}{A_1} = \frac{\varphi u (-\beta_u
+\beta_{\varphi})}{q}\right\vert_{x=-a}:= \varphi(-a)D_2,
\end{gather*}
where $-\beta_f$ stands for the logarithmic derivative of the
function $f(x)$. Hence we have
\begin{gather}
\psi_{\rm II}(x)=A_1 \varphi(-a) [C_2 \sin qx +D_2 \cos qx].
\label{psi2}
\end{gather}
In turn, from the condition of continuity at $x=a$, one arrives at
the expressions
\begin{gather*}
\frac{B_3}{A_1} = \varphi(-a) \frac{i}{2} {\cal F} (k,a), \qquad
\frac{A_3}{A_1} = \left(\frac{B_3}{A_1} \right)^*.
\end{gather*}
Here $z^*$ is the complex conjugate of $z\in \mathbb C$, and the
Jost function ${\cal F}(k,a)$ is def\/ined as
\begin{gather}
{\cal F}(k,a)=-\left\{\frac{e^{ikx}}{k} \left[u (\beta_- +
\beta_u)C_2 + v (\beta_- + \beta_v)D_2 \right] \right\}_{x=a},
\label{jost}
\end{gather}
with $-\beta_-$ the logarithmic derivative of $e^{-ikx}$. Thus,
for the solution in region III we have
\begin{gather}
\psi_{\rm III}(x)=  A_1\varphi(-a) \frac{i}{2} [{\cal F}(k,a)
e^{-ikx}- {\cal F}^*(k,a) e^{ikx}].
\label{psi3}
\end{gather}

{\bf Scattering states.} Consider now a particle of energy
$E=k^2>0$, arriving from $+\infty$ towards the ef\/fective zone of
the semi-harmonic well (\ref{pot}). The ref\/lection amplitude
\begin{gather}
s(k,a)=\frac{{\cal F}^*(k,a)}{{\cal F}(k,a)}
\label{s}
\end{gather}
is directly obtained from (\ref{psi3}) and has modulus one. This
can be written as $s(k,a) = e^{2i \delta(k,a)}$. We see that the
ef\/fect of the semi-harmonic rectangular well is to cause a phase
shift (up to integer multiples of $\pi$) of the ref\/lected wave by
\begin{gather}
\delta(k,a)=-\arctan\left[ \frac{\mbox{\rm Im}\,{\cal
F}(k,a)}{\mbox{\rm Re}\,{\cal F}(k,a)} \right].
\label{delta}
\end{gather}
Then, for scattering states, the solution of (\ref{schro}) cancels
as $x\rightarrow -\infty$, oscillates harmonically in region II
(see equation~\eqref{psi2}), and behaves as
\begin{gather*}
\psi_{\rm III}(x)= A_1 \varphi(-a) \vert {\cal F}(k,a) \vert
\sin(kx+\delta)
\end{gather*}
in region III.

\subsection{Conditions for the trapping of particles}
\label{sec:2.1}

The solutions of the Schr\"odinger equation (\ref{schro}) which
correspond to either bound or resonance states are picked out from
(\ref{psi1}), (\ref{psi2}) and (\ref{psi3}) such that only
outgoing waves exist. In other words, to construct these
solutions, it must be imposed the Siegert condition
\begin{gather}
\lim_{x\rightarrow +\infty} \beta_{\psi}(x) = -ik, \qquad k\in
\mathbb C.
\label{siegert}
\end{gather}
This last expression is equivalent to make ${\cal F}(k,a)=0$ in
(\ref{psi3}). Then, the Jost function (\ref{jost}) is analytically
continued so that the ref\/lection amplitude (\ref{s}) is a
meromorphic function with poles on the zeros of ${\cal F}(k,a)$.
For simplicity, hereafter $A_1=1/\varphi(-a)$. Therefore, the wave
functions of either bound states or resonances are written as
\begin{gather}
\psi(x)=\left\{
\begin{array}{ll}
\displaystyle\frac{\varphi(x)}{\varphi(-a)}, & x\leq -a,\\[2.0ex]
C_2 \sin qx + D_2 \cos qx, & \vert x\vert <a, \\[2.0ex]
-\frac{i}{2} {\cal F}^*(k,a) e^{ikx}, & a\leq x,
\end{array}
\right.  \qquad \mbox{such that} \quad {\cal F}(k,a)=0.
\label{sol}
\end{gather}
Given $a\neq 0$, the zeros of the Jost function ${\cal F}(k,a)$
are def\/ined by the roots of the transcendental equation
\begin{gather}
\beta_{\varphi}(-a)=-ik + \left[ \frac{ik \beta_{\varphi}(-a)-q^2}{q}
\right] \tan 2qa.
\label{roots}
\end{gather}
It is straightforward to verify that this last equation has no
solutions on the real line.

\begin{table}[t]
\centering
\caption{The dimensionless single bound state energy of a unit
area rectangular well in the free-particle (FP) and semi-harmonic
(SH) backgrounds for dif\/ferent values of $a$. In both cases, the
numerical results include the delta well as a limit ($a\rightarrow
0$).}\label{Tab1}

\begin{tabular}{|c|c|c|}
\hline\hline
$a$ & rectangular well & rectangular well\\
    &    (FP)     &  (SH)\\
\hline
  2.0 & $-0.113438$ & $-0.045272$ \\
  1.5 & $-0.130400$ & $-0.039482$ \\
  1.0 & $-0.153960$ & $-0.033514$ \\
  0.5 & $-0.189338$ & $-0.037435$ \\
  0.0 & $-0.25\phantom{0000}$ & $-0.079710$\\
\hline\hline
\end{tabular}

\end{table}

\subsubsection{Bound states}
\label{bound}

\begin{figure}[t]
\centering\includegraphics[width=6cm]{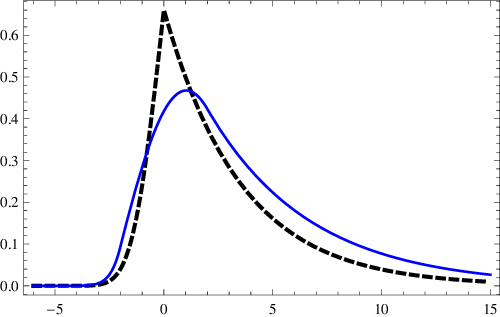}

\caption{The wave function of the unique bound state belonging to
the semi-harmonic rectangular well of unit area with $a=2$
(continuous, blue) and $a=0$ (dashed, black). The dimensionless
energies are $E_{a=2}=-0.045272$ and $E_{\delta}=-0.079710$
respectively (see Table~\ref{Tab1}). Notice the left-right
asymmetry produced by the parabolic part of potential~(\ref{pot}).}
\label{Fig:ground}
\end{figure}

Potential (\ref{pot}) includes a region of classical conf\/inement
in which some bound state energies could be present. In this
context, if $k=i\kappa$ is a zero of the Jost function with
$\kappa>0$, then $E=k^2=-\kappa^2$ is the energy eigenvalue of a
bound state. Indeed, if $k$ is a solution of (\ref{roots}) on the
positive imaginary axis of the complex $k$-plane, one has $\alpha=
\frac{1-k^2}{4} = \frac{1+\kappa^2}{4}> 0$, and $\beta_{\varphi}
\in \mathbb R$. The number $N$ of these roots is determined by the
area $A=2aV_0$ of the well. Thereby, the bound states are
represented by the square-integrable functions
\begin{gather*}
\psi_n(x)=\left\{
\begin{array}{ll}
\frac{\varphi_n(x)}{\varphi_n(-a)},  & x\leq -a,\\[2ex]
C_2 \sin q_n x + D_2 \cos q_n x, & \vert x\vert <a, \\[2ex]
\left. \dfrac{u(\beta_u+\kappa_n)C_2
+v(\beta_v+\kappa_n)D_2}{2\kappa_n}\right\vert_{x=a} e^{-\kappa_n
(x-a)}, & a\leq x,
\end{array}
\right. 
\\  n=0,1,\ldots, N(A)-1.
\nonumber
\end{gather*}
For instance, if $V_0^{-1} = 2a$, potential (\ref{pot}) becomes a
unit area rectangular well in a semi-harmonic background. This
admits a unique bound state, just as it is reported on
Table~\ref{Tab1} for f\/ive dif\/ferent values of $a$ (see also
Fig.~\ref{Fig:ground}). For comparison, the equivalent results
for a free-particle background are also reported. In both cases,
the single bound energy goes to a def\/inite negative number as
$a\rightarrow 0$. However, given $a\geq 0$, this energy is less
negative in the semi-harmonic background than in the free-particle
one. The same ef\/fect is observed for all the bound states
belonging to a rectangular well of arbitrary area $A \neq 0$.
Thus, in a semi-harmonic background, the bound states of the
rectangular well are displaced towards the threshold. Notice that
the oscillator-like interaction to the left of a given (symmetric)
well produces a left-right asymmetry in the involved wave
functions (see, e.g., Fig.~\ref{Fig:ground}).

\subsubsection{Resonances}
\label{resonances}

\begin{figure}[t]
\centering{\includegraphics[width=5cm]{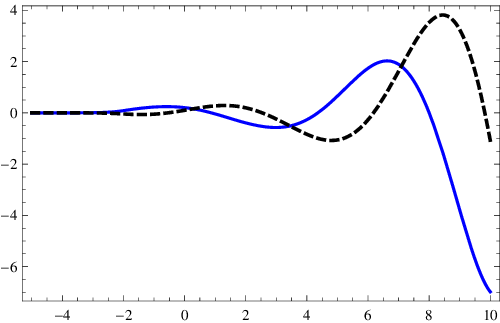} \hskip1cm
\includegraphics[width=5cm]{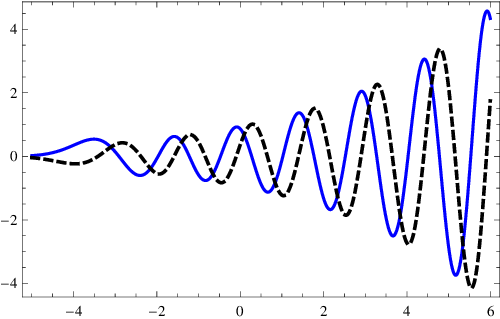}}

\caption{The real (continuous, blue) and imaginary (dashed, black)
parts of the Siegert function belonging to the f\/irst (left) and
f\/ifth (right) resonances reported on Table~\ref{Tab2} for $a=2$.
In all cases, the left-right asymmetry is due to the parabolic
part of potential (\ref{pot}).}
\label{Fig:gamow}
\end{figure}

\looseness=-1
Siegert functions are solutions of the Schr\"odinger equation
(\ref{schro}), associated to complex eigenvalues $\epsilon =
E-i\frac{\Gamma}{2}$, and fulf\/illing the purely outgoing condition
(\ref{siegert}). It is usual to consider $\epsilon$ as a~compound
of the resonance position $E$ and the inverse of the involved
lifetime $\tau^{-1}=\Gamma$. We obtain these complex eigenvalues
from the wave numbers~$k$ in the fourth quadrant of the complex
$k$-plane which are solutions of (\ref{roots}). Some of the f\/irst
resonances associated to a semi-harmonic rectangular well of unit
area are reported on Table~\ref{Tab2} for dif\/ferent values of $a$.
In contraposition to their equivalents in a free-particle
background, where the symmetry of the rectangular well is
inherited to the wave functions and resonances (see e.g.~\cite{Zav04} and \cite{Fer08b}), the Siegert functions of the
rectangular well in a semi-harmonic background cancel at
$x=-\infty$, obeying the left-right asymmetry generated by the
oscillator-like interaction. This is illustrated in
Fig.~\ref{Fig:gamow}.

\begin{table}
\centering

\caption{The f\/irst f\/ive resonances $\epsilon=E-i\frac{\Gamma}{2}$
belonging to the unit area rectangular well in a semi-harmonic
background for dif\/ferent values of $a$. The case $a=0$ (the
semi-harmonic delta well) is such that $\mbox{Re}(\epsilon)=
4m+3+\gamma_m$, with $\gamma_m \lessapprox 1$ and $m=0,1,\ldots$
(see Fig.~\ref{Fig:Deltares}).}\label{Tab2}

\vspace{1mm}

\begin{tabular}{|c|c|c|c|}
\hline\hline
$a$ & resonances & $a$ & resonances\\
\hline
 & $00.623117-i0.599545$ &  & $03.569260-i1.487849$\\
& $05.260402-i2.255076$ &  & $10.233701-i2.499541$\\
2 & $08.875649-i2.603548$ & 0.5 & $13.960171-i 2.280915$\\
& $11.217715-i1.995880$  &  & $17.904306-i 2.258438$\\
& $17.359977-i 2.229026$ &  & $21.878702-i 2.356575$\\
\hline
& $01.009578 -i0.981433$ &  & $03.792859-i 0.909297$\\
& $03.852457-i 2.106492$ &  & $07.852027-i 1.117703$\\
1.5 & $07.574481-i 2.017753$ & $5\times 10^{-4}$ & $11.880118 -i 1.242053$\\
& $09.826328-i 2.651538$ &  & $15.897157 -i 1.331170$\\
& $13.723866-i 2.196828$ &  & $19.908829 -i 1.400707$\\
\hline
& $01.838241-i 1.632446$ &  & $03.792839 -i 0.909196$\\
& $05.675163-i 1.760804$ &  & $07.852012 -i 1.117599$\\
1 & $08.749583-i 2.720863$  & 0 & $11.880106-i 1.241947$ \\
& $11.835116-i 2.294470$ &  & $15.897146-i 1.331064$\\
& $15.830298-i 2.343309$ &  & $19.908819 -i 1.400600$\\
\hline\hline
\end{tabular}
\end{table}

\section{The semi-harmonic delta well}
\label{sec:3}

\looseness=1
A particular case of the previous results which deserves special
attention is obtained in the limit $a\rightarrow 0$. It is well
known that the family of rectangular wells of unit area converges
to the delta well in the sense of distribution theory (see e.g.~\cite{Neg02} and~\cite{Chr03}). The transparency properties of the
point interactions are such that no resonances can be associated
to $\pm \Omega \delta(x)$, with $\Omega$ an opacity parameter
\cite{Flu99}. The situation changes if the free-particle
background is replaced by a~less trivial scenario. Namely, the
presence of resonances is ensured for the point interactions
either by putting a constant potential as the background
\cite{Lud87,Emm00} (see also \cite{Emm05}), or by constraining the
potential domain of $\delta (x-x_0)$ to be $[0,+\infty)$ rather
than the straight-line~$\mathbb R$~\cite{Flu99}. In
contradistinction, point dipole interactions represented by the
derivative of the delta distribution $\pm \delta'(x)$ admit
resonances in a natural form~\cite{Chr03}. We are going to analyze
the bound states and resonances associated to a delta well in the
semi-harmonic background introduced in the previous section.

\pagebreak

\begin{figure}[t]
\centering\includegraphics[width=5cm]{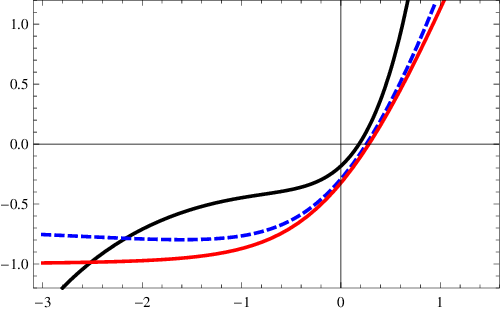}

\caption{The isolated zero of the Jost function which corresponds
to the unique bound state of a~semi-harmonic rectangular well of
unit area for $a=1$ (continuous, black), $a=0.1$ (dashed, blue)
and $a=0$ (continuous, red). In each case, the curve has been
depicted as a function of $k=i\kappa$, with $\kappa \in \mathbb
R$.}
\label{Fig:Jost}
\end{figure}

Potential (\ref{pot}) represents a family of functions of compact
support (the unit area rectangular wells, parameterized by $a$) in
a semi-harmonic background. If $a\rightarrow 0$, we have
\begin{gather*}
V_{\delta}(x):=\lim_{a\rightarrow 0} V(x;a)= V_s(x)-\delta(x),
\end{gather*}
with
\begin{gather*}
V_s(x)=\left\{
\begin{array}{ll}
x^2, & x<0,\\
0, & x>0
\end{array}
\right.
\end{gather*}
(details of the limit procedure for this kind of potentials can be
consulted in~\cite{Neg02,Neg04}). Applying the same limit to the
transcendental equation (\ref{roots}), one gets the expression
\begin{gather}
2\Gamma_{\mbox{e}} \left(\frac{3-k^2}{4}\right)=(1+ik)
\Gamma_{\mbox{e}} \left(\frac{1-k^2}{4}\right).
\label{droots}
\end{gather}
Equation (\ref{droots}) admits an isolated root on the positive
imaginary axis of the complex $k$-plane (see Fig.~\ref{Fig:Jost}).
We have $k_{\delta}= i0.2823302$. The corresponding energy
$E_{\delta}= k_{\delta}^2= -0.0797104$, is consistently recovered
as the limit $a\rightarrow 0$ of the values reported on
Table~\ref{Tab1}. Interestingly, this bound energy is closer to
zero than the bound energy of the delta well in a free-particle
background $E=-0.25$. That is, the semi-harmonic background causes
the displacement of the single bound energy towards the threshold,
just as it has been remarked above. The involved wave function
\begin{gather*}
\psi_{\delta}(x)=\left\{
\begin{array}{rl}
\varphi(x), & x\leq 0,\\
e^{-\kappa x}, & 0<x
\end{array}
\right.
\end{gather*}
is depicted in Fig.~\ref{Fig:ground}. Observe that the parabolic
part of the potential produces a decreasing of~$\varphi(x)$ which
is faster than the decreasing of its counterpart $e^{-\kappa x}$
in the free particle zone.

\begin{figure}[t]
\centering{\includegraphics[width=8cm]{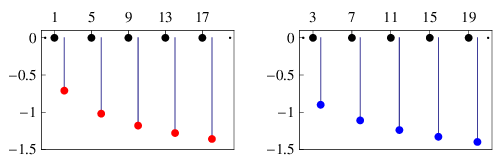}}

\caption{First f\/ive resonances $\epsilon = E-i\frac{\Gamma}{2}$ of
a delta barrier (left, red disks) and a delta well (right, blue
disks) in a semi-harmonic background. Black disks on the real line
represent the even and odd energies of the harmonic oscillator
respectively. Vertical lines are included as a reference. The
numerical values of $\epsilon$ can be consulted on Tables
\ref{Tab3} and \ref{Tab2} respectively.}
\label{Fig:Deltares}
\end{figure}

In the present case, the resonances are determined by the roots of~(\ref{droots}) in the fourth quadrant of the complex $k$-plane. Up
to a constant $\gamma_m$, the position $E_m$ of each of these
`complex energies' is clearly connected to the distribution of the
odd energy eigenvalues of the harmonic oscillator:
$\mbox{Re}(\epsilon) = 4m + 3 + \gamma_m$, with $m =0,1,2,\ldots$.
For the results reported on Table~\ref{Tab2}, the para\-me\-ter~$\gamma_m$ approaches~1 from below as $m$ increases (see
Fig.~\ref{Fig:Deltares}). As a conclusion, dif\/ferent than the
conventional delta well, a delta well in a semi-harmonic
background admits resonances. More details will be given in
Section~\ref{sec:5}.

\section{Semi-harmonic barriers}
\label{sec:4}

\begin{figure}[t]
\centering{\includegraphics[width=5cm]{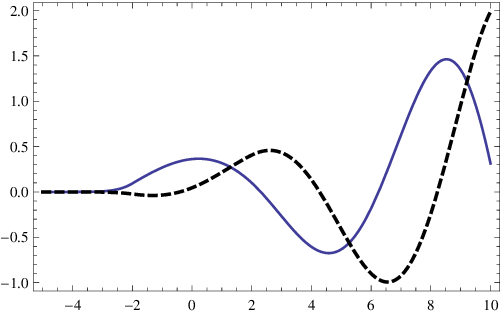} \hskip1cm
\includegraphics[width=5cm]{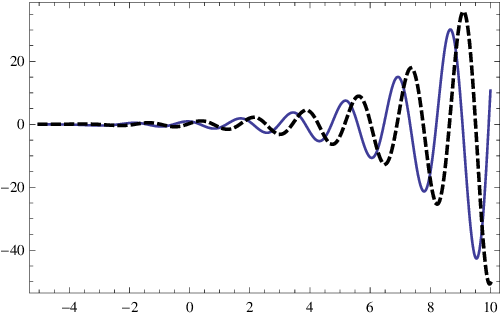}}

\caption{The real (continuous, blue) and imaginary (dashed, black)
parts of the Siegert function belonging to the f\/irst (left) and
f\/ifth (right) resonances reported on Table~\ref{Tab3} for $a=2$.
Remark that the left-right asymmetry of all these functions is
generated by the parabolic part of potential~(\ref{pot}).}
\label{Fig:Bgamow}
\end{figure}

The case of a particle in a semi-harmonic rectangular barrier can
be analyzed as a consequence of the previous results. The change
$V_0 \rightarrow -V_0$ in (\ref{pot}), (\ref{schro}) produces $q
\rightarrow i \sqrt{\vert k^2 -V_0 \vert}$ for $k^2<V_0$, and $q
\rightarrow \sqrt{k^2 -V_0}$ for $k^2>V_0$. The straightforward
calculation shows that there is no bound states for this system.
Some of the corresponding resonances, in turn, are reported on
Table~\ref{Tab3} (see also Fig.~\ref{Fig:Bgamow}). In this case,
the complex energies $\epsilon$ are below and close to the
positive real axis such that, in the very limit $a\rightarrow 0$,
$\mbox{Re}(\epsilon)$ is in one of the positions $4m+1+\lambda_m$,
with~$\lambda_m$ a~constant and $m=0,1,2,\ldots$. For the results
on Table~\ref{Tab3}, $\lambda_m$ approaches 1 from above as $m$
increases (see Fig.~\ref{Fig:Deltares}). Then, the resonances of a
delta barrier in a semi-harmonic background are distributed in
correspondence with the even energy eigenvalues of the harmonic
oscillator. See more details in Section~\ref{sec:5}.

\begin{table}
\centering
\caption{The f\/irst f\/ive resonances $\epsilon=E-i\frac{\Gamma}{2}$
belonging to the unit area rectangular barrier in a semi-harmonic
background for dif\/ferent values of $a$. The case $a=0$ (the
semi-harmonic delta barrier) is such that $\mbox{Re}(\epsilon) =
4m+1+\lambda_m$, with $\lambda_m \gtrapprox 1$ and
$m=0,1,2,\ldots$ (see Fig.~\ref{Fig:Deltares}).}\label{Tab3}
\vspace{1mm}

\begin{tabular}{|c|c|c|c|}
\hline\hline
$a$ & resonances & $a$ & resonances\\
\hline
 & $00.595222-i0.312336$ &  & $01.957470-i0.802720$\\
& $04.211813-i1.853317$ &  & $05.720407-i1.561124$\\
2 & $07.387244-i2.485184$ & 0.5 & $09.297545-i2.218137$\\
& $10.572524-i1.985917$  &  & $12.626496-i2.505935$\\
& $12.865432-i2.874092$ &  & $16.213699-i2.433756$\\
\hline
& $00.846017-i0.497624$ &  & $02.076264-i0.718026$\\
& $06.188284-i2.213815$ &  & $06.065577-i1.025854$\\
1.5 & $08.833155-i2.011628$ & $5\times 10^{-4}$ & $10.058023-i1.181640$\\
& $12.476467-i2.723854$ &  & $14.052696-i1.286283$\\
& $14.938732-i2.618887$ &  & $18.048705-i1.360061$\\
\hline
& $01.313699-i0.776043$ &  & $02.076211-i0.718123$\\
& $04.317689-i1.833635$ &  & $06.065549-i1.025956$\\
1 & $07.338936-i2.031873$  & 0 & $10.058003-i1.181793$\\
& $10.901777-i2.284989$ &  & $14.052679-i1.286389$\\
& $14.229903-i2.730174$ &  & $18.048690-i1.365109$\\
\hline\hline
\end{tabular}
\end{table}

\subsection{Darboux transformations}

\begin{figure}[t]
\centering{\includegraphics[width=5cm]{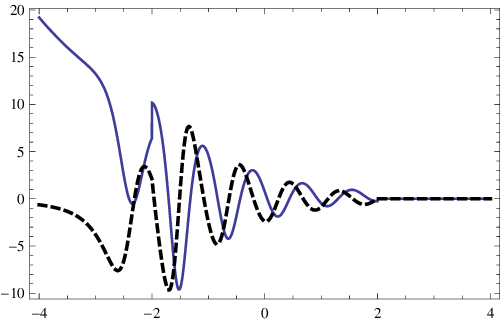} \hskip1cm
\includegraphics[width=4cm]{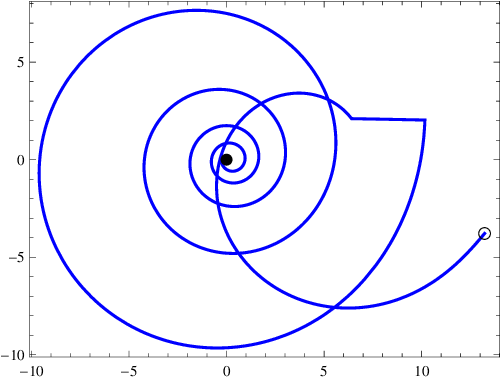}}

\caption{{\it Left.} The real (continuous, blue) and imaginary
(dashed, black) parts of the supersymmetric partner $\widetilde
V(x;a)$ of a semi-harmonic rectangular barrier. This has been
constructed with the f\/ifth resonance for $a=2$. {\it Right.} The
Argand diagram of $\widetilde V(x;2)$. The circle $\circ$ and disk
$\bullet$ serve as reference and correspond to $\widetilde
V(-3;2)$ and $\widetilde V(3;2)$ respectively.}
\label{Fig:susy}
\end{figure}

The Siegert condition (\ref{siegert}) is appropriate to construct
the complex supersymmetric partners of the semi-harmonic
potentials analyzed in the previous sections. Consider the Darboux
transformation
\begin{gather}
\widetilde V(x;a) = V(x;a) +2\beta'(x),
\label{darboux}
\end{gather}
with $V(x;a)$ either a semi-harmonic well or barrier, and $\beta
(x)$ the logarithmic derivative of any of the corresponding
Siegert functions $\psi_{\epsilon}$. Condition (\ref{siegert})
indicates that the new potential $\widetilde V(x;a)$ behaves as
$V(x;a)$ in the limit $x\rightarrow +\infty$. On the other hand,
the straightforward calculation shows that $\beta(x) \rightarrow
0$, as $x\rightarrow -\infty$. Therefore, $\widetilde V(x;a)$ is a
complex potential behaving as $V(x;a)$ at the edges of $\mbox{Dom}\,
V(x;a) = (-\infty, +\infty)$. The supersymmetric formalism of
quantum mechanics~\cite{And84,Mie04,Kha04,Bay04,And04,Ferd10}
ensures that the energy spectrum of $\widetilde V(x;a)$ is the
same as that of $V(x,a)$ if the superpotential is def\/ined as
$\beta(x) = -\frac{d}{dx} \ln \psi_{\epsilon}(x)$ (cf.~\cite{Fer08b,And99,Fer03,Ros03a,Ros07}). In Fig.~\ref{Fig:susy}
we show the behavior of the Darboux-deformed semi-harmonic
rectangular barrier, constructed from~(\ref{darboux}) with the
Siegert state belonging to the f\/ifth resonance of
Table~\ref{Tab3}. The presence of maxima and minima, in both the
real and imaginary parts of the potential, makes $\widetilde
V(x;a)$ to behave as an optical device which both refracts and
absorbs light waves \cite{Mie04,Fer08b}. A similar result is
obtained for the Darboux-deformations of the semi-harmonic wells.
The procedure can be repeated at will, giving rise to more
elaborate complex deformations of the initial potentials.
Particularly interesting, the double transformation obtained by
using a complex eigenvalue $\epsilon$ in a f\/irst step, and its
complex conjugate $\epsilon^*$ in the second step, produces
deformations which are real functions (see, e.g.~\cite{Fer03,Fer08b,Fer10}). In such a case, the semi-harmonic
barriers will exhibit `hair' as a characteristic of the
deformation~\cite{Fer08b,Fer10}.

 \section{Dwell time}
\label{sec:5}

\begin{figure}[t]
\centering\includegraphics[width=7cm]{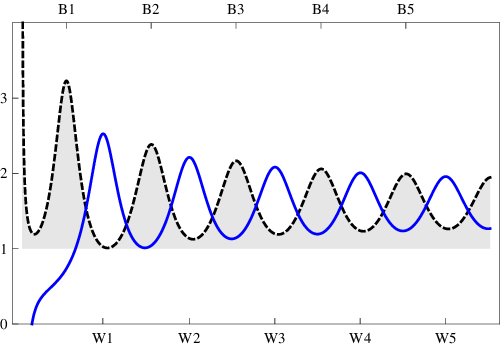}

\caption{The Eisenbud--Wigner time delay $\tau_W$ for the
scattering process of a semi-harmonic delta barrier (dashed,
black) and well (continuous, blue). The peaks W$n$ (B$n$) localize
the real part of the resonances reported on Table~\ref{Tab2}
(Table~\ref{Tab3}).}
\label{Fig:wigner}
\end{figure}

As yet the bound states and resonances have been analyzed as
dif\/ferent manifestations of the same sort of mathematical
solution. Both of them are def\/ined in terms of the zeros of the
Jost function (\ref{jost}) and are represented by a Siegert state
of the form (\ref{sol}). Concerning scattering states, let $E>0$
be the energy of a particle which impinges on a scatterer from the
right, according to the rule~(\ref{pot}). Here $V_0$ can be either
positive or negative. Exterior to the interaction zone $(-\infty,
a)$, the stationary scattering wave function is $\psi_{\rm III}
\propto e^{-ikx}-s(k,a)e^{ikx}$ (see equations~\eqref{psi3} and~\eqref{s}).
Since there is neither sources nor sinks we have total probability
conservation $\vert s(k,a) \vert^2=1$. We want to get some
insights on the time spent by the incident particle in the
interaction zone. To analyze the `capturing' of the particle by
the semi-harmonic rectangular potential, we assume $\vert s(k,a)
\vert^2 = e^{-\tau^a_D/\tau_f}$, with $\tau^a_D$ the mean time
spent by the particle in the interaction zone $(-\infty,a)$, and
$\tau_f$ a characteristic time constant to be determined. Using
the absorption probabilities method \cite{Gol90,Hua91,Mug92}, a
small imaginary part $\Delta V_I$ is added to the potential under
consideration. The absorption dimensionless Schr\"odinger equation
to be solved is $(H+i \Delta V_I)\psi_a = i\frac{\partial
\psi_a}{\partial t}$. Given a solution $\psi_0$ of the unperturbed
equation (for which $\Delta V_I=0$), we use the ansatz $\psi_a=
\psi_0 g$ to get $\psi_a = \psi_0 e^{\Delta V_I t}$ (an arbitrary
integration constant has been omitted for simplicity). Therefore,
the probability density of f\/inding the particle is $\vert \psi_0
\vert^2 e^{2\Delta V_I t}$. The time $\tau_f$ required for the
probability density to decrease a factor $e^{-1}$ of its initial
value $\vert \psi_0 \vert^2$ is $\tau_f = -1/(2\Delta V_I)$, as
this is introduced in \cite{Hua91}. The dwell time (or sojourn
time) $\tau^a_D$ in the region $(-\infty,a)$ can be now obtained
from $\vert s(k,a)\vert^2 = e^{2\tau^a_D \Delta V_I}$. Indeed, in
the limit of small~$\Delta V_I$, a simple calculation gives
\begin{gather}
\tau^a_D =\frac12 \lim_{\Delta V_I \rightarrow 0}
\frac{\partial}{\partial \Delta V_I} \vert s(k,a) \vert^2.
\label{HW}
\end{gather}
Now, consider a solution of the absorption eigenvalue equation $(H
+ i \Delta V_I -\epsilon) \psi_{\epsilon} = 0$, with $\epsilon \in
\mathbb C$. The derivative with respect to $\Delta V_I$ of this
last equation, after evaluating it for $\Delta V_I =0$, is reduced
to
\[
i\psi_{\epsilon} +(H-\epsilon) \left. \frac{\partial
\psi_{\epsilon}}{\partial \Delta V_I} \right\vert_{\Delta V_I
=0}=0.
\]
The derivative of the absorption eigenvalue equation with respect
to $\epsilon_I$, the imaginary part of~$\epsilon$, after
evaluating it for $\Delta V_I =0$, produces
\[
-i\psi_{\epsilon} +(H-\epsilon) \frac{\partial
\psi_{\epsilon}}{\partial \epsilon_I} =0.
\]
Adding these last two results yields
\[
\lim_{\Delta V_I \rightarrow 0} \frac{\partial}{\partial \Delta
V_I} = - \frac{\partial}{\partial \epsilon_I}.
\]
In this way, equation (\ref{HW}) can be rewritten as follows
\begin{gather}
\tau^a_D = -\frac12 \frac{\partial}{\partial \epsilon_I} \vert
s(k,a) \vert^2.
\label{HW2}
\end{gather}
Since $k^2=\epsilon$, the ref\/lection amplitude $s(k,a)$ must be a
complex analytic function of the complex eigenvalue $\epsilon = E
+i\epsilon_I$. The Cauchy--Riemann condition reads as
\[
\frac{\partial s_R}{\partial E} = \frac{\partial s_I}{\partial
\epsilon_I} \qquad \mbox{and} \qquad -\frac{\partial s_R}{\partial
\epsilon_I} = \frac{\partial s_I}{\partial E}.
\]
The introduction of these last expressions into (\ref{HW2})
produces
\begin{gather}
\tau^a_D = 2\frac{\partial \delta}{\partial E} = \tau_W,
\label{relation}
\end{gather}
where we have used $s(k,a)=e^{2i\delta(k,a)}$. From the above
expression, the dwell time of the scattering particles which are
`trapped' in the semi-harmonic rectangular potentials is in direct
connection with the slope of the ref\/lected wave phase shift
$\delta(E,a)$. Thereby, the dwell time~(\ref{relation}) coincides
with the def\/inition of phase time $\tau_W$, introduced by Eisenbud
and Wigner as the time delay in binary collisions~\cite{Wig55}. In
particular, the formation of a resonance $\epsilon =E_r -i
\frac{\Gamma}{2}$ introduces a~positive time delay between the
arrival and departure of the scattering particle from the region
$(-\infty,a)$. This last means that particles impinging the
semi-harmonic rectangular potentials with energy $E=E_r$, will
spend times in~$(-\infty, a)$ which are larger than the times
spent by the scattering particles of energy $E=E_r+\varepsilon$,
with $\varepsilon \neq 0$ a real number small enough. A rapid
increasing of the phase shift is then expected in the vicinity of
the resonance position~$E_r$. Hence, the phase shift~(\ref{delta})
encodes enough information to identify true resonances by
associating the peaks of~(\ref{relation}) with the real part of
the complex eigenvalues~$\epsilon$ (see, e.g.~\cite{Tay06}). Of
special interest, the time delay of point interactions in a
semi-harmonic background can be calculated as the limit
$a\rightarrow 0$ of~(\ref{relation}). One obtains
\begin{gather}
\tau_W^{(\pm)} (E) =- 2 \left(\frac{1}{1+W^2_{\pm}(E)}\right)
\frac{dW_{\pm}(E)}{dE}
\label{wigner}
\end{gather}
where the sign $+$ ($-$) stands for the semi-harmonic delta
barrier (well) and
\begin{gather*}
W_{\pm}(E) =\frac{\mbox{Im}(\cal F_{\pm})}{\mbox{Re}(\cal
F_{\pm})} =\frac{- \Gamma_{\mbox{e}} \left(\frac{1-E}{4}\right)
\sqrt{E}}{\pm \Gamma_{\mbox{e}} \left(\frac{1-E}{4} \right) + 2
\Gamma_{\mbox{e}} \left( \frac{3-E}{4} \right)}, \qquad {\cal
F}_{\pm}:= {\cal F}_{\pm} (x, a\rightarrow 0).
\end{gather*}
An straightforward calculation shows that $\tau_W^{(\pm)}
\rightarrow \frac{\pi}{2}$, as $E \rightarrow +\infty$. From
Fig.~\ref{Fig:wigner}, we notice that $\tau_W^{(-)}$
$\big(\tau_W^{(+)} \big)$ exhibits a series of
Fock--Breit--Wigner peaks W1, W2, $\ldots$ (B1, B2, $\ldots$),
centered each one at the real part of the f\/irst, second, etc,
energy resonances reported on Tab\-le~\ref{Tab2} (Tab\-le~\ref{Tab3}).
Thus, the zeros of the Jost function~(\ref{jost}) are interrelated
with the maxima of the time delay~(\ref{wigner}), as expected for
true resonances~\cite{Tay06}. A conclusion which makes
self-consistent our approach since the same results are obtained
in two dif\/ferent forms. From Fig.~\ref{Fig:wigner}, we also
notice that all the `captured' particles spend a time $\tau_W \geq
1$ in $(-\infty, a)$, except for a region of small positive
energies in the rectangular well case. The closer to the resonance
position $\mbox{Re}(\epsilon)$ is the incoming energy $E$, the
larger is the time spent by the particle in the interaction zone.

In general, after entering the region $(-\infty,a)$ of an
arbitrary semi-harmonic rectangular potential, the particle may be
ref\/lected back and forth many times before it manages to get back
out (see discussions on the matter in \cite{Boh51}). These
ref\/lections may take place either in the zone of the rectangular
potential $(-a,a)$, the parabolic part of the potential $(-\infty,
-a)$, or in a combination of both. Our calculations consider the
complete scattering process so that it is not possible to
distinguish among all possible internal ref\/lections and
self-interference during the approach to the region $(-\infty,
a)$. For an arbitrary isotropic potential $V_0$ (i.e., $V_0(x)$
becomes zero as $\vert x \vert \rightarrow +\infty$), the dwell
time $\tau_D$ must obey the identity $\tau_D = \tau_R \vert R
\vert^2 + \tau_T \vert T \vert^2$, with $R(T)$ the ref\/lection
(transmission) amplitude, and $\tau_R$ ($\tau_T$) the average time
spent in $V_0$ by ref\/lected (transmitted) particles. For instance,
if $V_0$ is a rectangular barrier in the free-particle background,
one has $\tau_R=\tau_T = \tau_D$, so that all particles spend, on
average, the same amount of time in $V_0$, no matter the size of
the barrier \cite{Jaw88,Hau89,Gol90}. Our expression (\ref{HW2})
is then associated to the ref\/lection time, the mean time if
particles are f\/inally ref\/lected, in the absorption model
\cite{Hua91} (see also \cite{Jaw88}). The comparison of
$\tau^a_D$, as this has been def\/ined in (\ref{relation}), with
$\tau_R$ for the rectangular barrier could give us additional
information of the time delay induced by the semi-harmonic
background.

Roughly speaking, time delay of the semi-harmonic rectangular
potentials should correspond to the excess time that the
scattering particle spends in $(-\infty, a)$, when compared to a
free particle under the same initial conditions. Since our model
considers the semi-harmonic background as the {\em environment}
into which the square potential is embedded, this last can be
taken as an open quantum system \cite{Rot09}. In this respect, the
above indicated comparison of $\tau^a_D$ and $\tau_R$ makes sense
in order to calculate time delays as the dif\/ference of ref\/lection
times. Some insights would be also obtained concerning
transmission times. Given an open interval $\omega =(x_1,x_2)$ of
suf\/f\/iciently large size $x_2 -x_1$ in $(-\infty, a)$, and a time
window $\theta = (t_1,t_2)$ with $t_2>t_1$, one could introduce
the corresponding dwell time as
\[
\Theta(\omega, \theta) = \int_{\theta} dt \int_{\omega} dx \vert
e^{-iHt} \psi(x) \vert^2,
\]
with $H$ as this has been given in (\ref{schro}), and $\psi$ an
scattering state of energy $E$. In a similar form, the expression
\[
\Theta_0(\omega, \theta) = \int_{\theta} dt \int_{\omega} dx \vert
e^{-iH_0t} \psi(x) \vert^2,
\]
with $H_0$ the Hamiltonian of the square potential in a
free-particle background, could be taken as the dwell time of the
`free-particle' system. The time delay in the space-temporal
window $(\omega, \theta)$ should correspond to the dif\/ference
$\Theta- \Theta_0$, as usual. In such a def\/inition the quantities~$\Theta$ and~$\Theta_0$ are assumed to be well def\/ined and f\/inite
for arbitrary $\omega \in (-\infty,a)$. In principle, it seems to
be the case since all the involved scattering wave functions
cancel at $x=-\infty$ (this is because the parabolic part of the
environment). In order to arrive at a quantity which is
independent of $\omega$, one could take the limits $x_1
\rightarrow -\infty$ and $x_2 \rightarrow a$. However, the
anisotropy of the semi-harmonic rectangular potentials makes this
last step not evident a priori. That is, some caution is required
to apply the dif\/ference of sojourn times as the def\/inition of time
delay in the semi-harmonic rectangular potentials. Mainly, as
connected with transmission dwell times in the interaction zone
$(-\infty,a)$ since the potential diverges as $x^2$ at
$x=-\infty$. Thus, in this last case, there is not a clear
connection between sojourn times and phase times because almost
all the approaches on the matter require potentials localized in a
f\/inite zone of the real line. The problem deserves a detailed
analysis and the results will be reported elsewhere.

\section{Concluding remarks}
\label{sec:6}

In this paper, we studied the energy properties of a particle in
the presence of a semi-harmonic rectangular potential. The latter
is a one-dimensional rectangular potential in a background
composed by an harmonic interaction to the left and a free
particle interaction to the right of the rectangular potential.
The advantage of this model lies in its simplicity to get analytic
expressions for the wave functions, scattering states and
resonances. Indeed, the problem is faced by considering bound
states and resonances as dif\/ferent manifestations of the same sort
of mathematical solution. Both of them are def\/ined in terms of the
zeros of the Jost function and are represented by a Siegert state.
The numerical integration of the energies is done by solving a
simple transcendental equation. In particular, it is found that,
in the presence of a semi-harmonic background, the conventional
bound energies of the rectangular well are displaced towards the
threshold. Such behavior is preserved in the limit where the well
becomes a delta. In this limit, there is a single bound state of
energy $E_{\delta}=-0.079710$, which is less negative than its
counterpart in a free-particle background, the value of which is
$E=-0.25$. The resonances appearing for the semi-harmonic delta
well exhibit a very peculiar behavior. They are distributed below
and close to the positive real axis of the complex plane in such a
way that their real part mimics the odd energy eigenvalues
distribution of a harmonic oscillator. Namely, they are located
according to the rule $4m+3+\gamma_m$, with $\gamma_m \lessapprox
1$ and $m=0,1,2,\ldots$. A similar situation occurs for the
resonances of a semi-harmonic delta barrier, with the distribution
ruled by $4m+1+\lambda_n$ with $\lambda_m \gtrapprox 1$. Thus, in
this case, the semi-harmonic background induces the real part of
the resonances to be distributed in correspondence to the energy
eigenvalues of the harmonic oscillator. The Siegert functions were
successfully applied to construct complex Darboux-deformations of
the semi-harmonic rectangular potentials. These new potentials
behave as optical devices which both refract and absorb light
waves.

\looseness=-1
Concerning the times involved in the scattering process, we
assumed the time exponential decay rule $\vert s(k,a) \vert^2 =
\exp(-\tau^a_D/\tau_f)$, with $s(k,a)$ the ref\/lection amplitude,
$\tau_f$ a constant to be determined and $\tau^a_D$ the mean time
spent by an scattering particle coming from $+\infty$ towards the
interaction zone $(-\infty, a)$. The application of the absorption
probability method \cite{Gol90,Hua91,Mug92} allowed us to show
that~$\tau^a_D$ coincides with the phase-time of Eisenbud and
Wigner~$\tau_W$, calculated as the derivative of the ref\/lected
wave phase shift~$\delta (E,a)$ with respect to the energy~$E$.
Becuase the global properties of the potentials we have discussed
on, the dwell time so calculated corresponds to a ref\/lection time
(the mean time if particles are f\/inally ref\/lected). In the limit
where the rectangular potentials become point-like interactions,
we derived an explicit analytical expression for $\tau_W$, and
showed that its local maxima are in correspondence with the
resonances obtained in the previous sections. Finally, the
semi-harmonic rectangular potentials can be viewed as the open
quantum systems integrated by a rectangular potential (the system
itself) and an environment (the semi-harmonic background). In this
picture, the system is an isotropic potential for which the time
spent in any f\/inite region of space, averaged over all incoming
particles, is well def\/ined and corresponds to the dwell time
$\tau_D$ involved. Therefore, the semi-harmonic rectangular
potentials can be seen as the `system' af\/fected by a semi-harmonic
interaction. Some insights on time delay can be obtained by
calculating dif\/ferences between the dwell time of these two
systems in a given space-temporal window. This approach, however,
is not directly applicable in the present case because the
anisotropy of the environment. Work in this direction is in
progress.

\appendix

\section{The conf\/luent hypergeometric operator}
\label{apA}

\begin{figure}[t]
\centering\includegraphics[width=5cm]{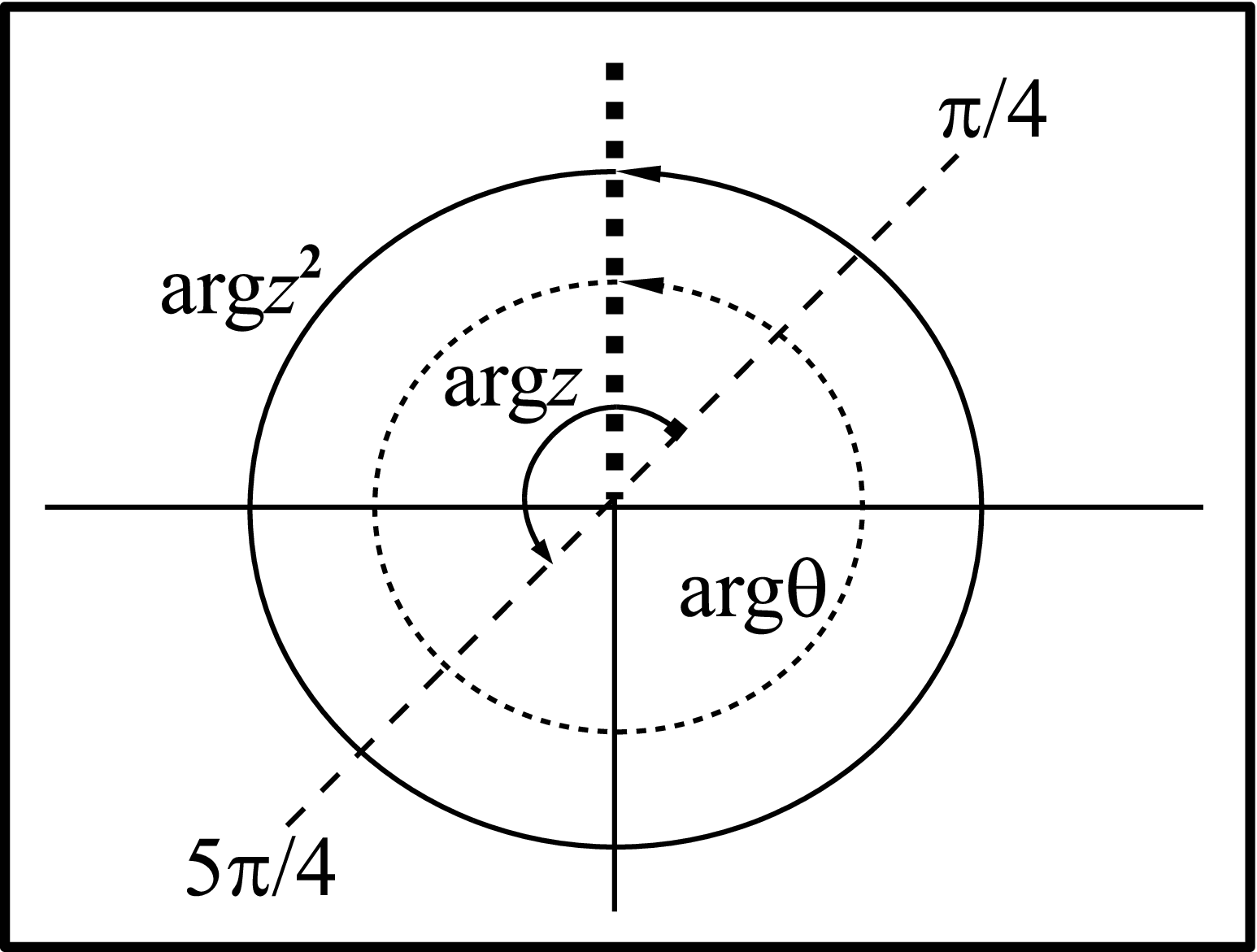}

\caption{The two f\/irst sheets of the Riemann surface def\/ined by
the factor $e^{-a\pi i}$ in equation~(\ref{asymp}), the cut is at
$\frac{\pi}{2}$. If $\frac{\pi}{4}<\arg z <\frac{5\pi}{4}$, then
$z^2$ and $\theta=e^{-2\pi i}z^2$ are in the second and f\/irst
sheets respectively.}
\label{Fig:app1}
\end{figure}

The kernel ${\cal K}_{(a,c)}$ of the conf\/luent hypergeometric
operator
\begin{gather*}
L_{(a,c)}:=z\frac{d^2}{dz^2} + (c-z)\frac{d}{dz} -a, \qquad a,c\in
\mathbb R, \qquad z\in \mathbb C
\end{gather*}
is integrated by the solutions of the Kummer equation $L_{(a,c)}
f(a,c;z)=0$. That is, if $f(a,c;z)$ is a conf\/luent hypergeometric
function then $f(a,c;z)\in {\cal K}_{(a,c)}$ \cite{Neg00,Ros03}.
The asymptotic expansion of $f(a,c;z)$ for large values of $\vert
z\vert$ is given by
\begin{gather}
f(a,c;z) = \displaystyle\frac{\Gamma(c)}{\Gamma(c-a)} e^{\pm a\pi
i} z^{-a} [1+P_-(a)]+ \displaystyle \frac{\Gamma(c)}{\Gamma(a)}
e^z z^{a-c}[1+P_+(c-a)]
\label{asymp}
\end{gather}
where the positive sign in the factor $e^{\pm a\pi i}$ is taken
when $-\frac{\pi}{2}<\arg z < \frac{3\pi}{2}$, while the negative
sign is taken when $-\frac{3\pi}{2} < \arg z <\frac{\pi}{2}$
\cite{Wan89} and
\[
P_{\pm}(\gamma)= \sum_{n=1}^{\infty} (\pm)^n \frac{(\gamma)_n(1-c+
\gamma )_n}{n!z^n}, \qquad (\gamma)_n=\gamma(
\gamma+1)(\gamma+2)\cdots (\gamma+n-1), \quad (\gamma)_0=1.
\]
The above expressions can be utilized to deduce the asymptotic
expansion of $f(a,c;z)$ for specif\/ic ranges of $\arg z$. For
instance, if $-\frac{\pi}{2}<\arg z<\frac{\pi}{2}$ we get
\begin{gather*}
f(a,c;z) \rightarrow   \frac{\Gamma(c)}{\Gamma(a)} e^z
z^{a-c}[1+P_+(c-a)], \qquad \mbox{\rm Re} (z)>0
\end{gather*}
and
\begin{gather}
f\big(a,c;z^2\big) \approx \displaystyle \frac{\Gamma(c)}{\Gamma(a)}
e^{z^2} z^{2(a-c)} \qquad \mbox{\rm if} \quad z\rightarrow
+\infty.
\label{2asymp+}
\end{gather}
To f\/ind the behavior of $f(a,c;z^2)$ for $z\rightarrow -\infty$ we
consider a range of $\arg z$ which includes the negative real
values of $z$. For this, let us assume $\frac{\pi}{4}<\arg
z<\frac{5\pi}{4}$ (see Fig.~\ref{Fig:app1}). Then
$\frac{\pi}{2}<\arg z^2<\frac{5\pi}{2}$, and $z^2$ is in the
second sheet of the Riemann surface def\/ined by the factor
$e^{-a\pi i}$ in~(\ref{asymp}). Since $\theta=e^{-2\pi i} z^2$ is
such that $-\frac{3\pi}{2}< \arg \theta <\frac{\pi}{2}$, from
(\ref{asymp}) we arrive at
\[
f(a,c;\theta)=  \frac{\Gamma(c)}{\Gamma(c-a)} e^{a\pi
i} z^{-2a} [1+P_-(a)]+   \frac{\Gamma(c)}{\Gamma(a)}
e^{z^2} z^{2(a-c)}e^{-2\pi i(a-c)}[1+P_+(c-a)].
\]
Therefore, if $\arg z=\pi$ one f\/inally gets
\begin{gather}
f\big(a,c;z^2\big) \approx   \frac{\Gamma(c)}{\Gamma(a)}
e^{z^2} z^{2(a-c)}e^{-2\pi i(a-c)}\qquad \mbox{\rm if} \quad
z\rightarrow -\infty.
\label{2asymp-}
\end{gather}

\subsection*{Acknowledgements}

The support of CONACyT and DGAPA (UNAM) is acknowledged.

\pdfbookmark[1]{References}{ref}
\LastPageEnding

\end{document}